\documentclass[preprint,onecolumn,nofootinbib]{revtex4}

\usepackage[colorlinks=true,linkcolor=blue,urlcolor=blue,filecolor=black,citecolor=red,pdfstartview=FitV,pdftitle={},pdfsubject={},pdfkeywords={},pdfpagemode=None,bookmarksopen=true]{hyperref}
\usepackage{graphicx}
\usepackage{amsmath}
\usepackage{amsfonts}
\usepackage{amssymb,ulem}
\usepackage{color}
\usepackage{tikz}
\usepackage{dcolumn}
\usepackage{xcolor}

\begin{document}
\title{
  Mixed State Entanglement For Holographic Systems With A Scalar Hair
}
\author{Zheng-Chen Song $^{1}$}
\email{creddd346@gmail.com}
\author{Zhe Yang $^{1}$}
\email{yzar55@stu2021.jnu.edu.cn}
\author{Chong-Ye Chen $^{1}$}
\email{cycheng@stu2018.jnu.edu.cn}
\author{Peng Liu $^{1}$}
\email{phylp@email.jnu.edu.cn}
\affiliation{
  $^1$ Department of Physics and Siyuan Laboratory, Jinan University, Guangzhou 510632, China
}

\begin{abstract}

  We study the mixed state entanglement of an asymptotic AdS black hole system with scalar hair. Through numerical calculations, we find that the holographic entanglement entropy (HEE) presents a non-monotonic behavior with the system parameter, depending on the size of the subregion. In addition, the mutual information (MI) also shows non-monotonic behavior in certain ranges of system parameters. However, the entanglement wedge minimum cross-section (EWCS), which is a mixed state entanglement measure, increases monotonically with the AdS radius; meanwhile, shows non-monotonic with the temperature. We also give analytical understandings of the phenomena above.

\end{abstract}
\maketitle
\tableofcontents

\section{Introduction}
\label{sec: introduction}

Quantum information (MI) has become increasingly important in recent years. Especially, MI has attracted heavy attention in many interdisciplinary fields, like quantum computing, quantum communication, and condensed matter theory (CMT) \cite{Osterloh:2002na, Amico:2007ag, Wen:2006topo, Kitaev:2006topo}. There are many forms of quantum information, among which the entanglement is the most significant property.

However, measurements of entanglement are difficult to compute in CMT, especially in a strongly correlated system. With the increase of the degree of freedom, the dimension of Hilbert space exponentially increases. AdS/CFT provides a novel tool to overcome this problem in strongly correlated systems. According to the AdS/CFT theory, a gauge field is dual to a gravity system \cite{Maldacena:1997,hooft:1993}. This duality makes the study of strongly correlated systems practicable and builds a bridge between a quantum system and gravity spacetime. Moreover, AdS/CFT offers intriguing paths to understand quantum information and the emergency of spacetime \cite{Ryu&Takayanagi:2006, Lewkowycz:2013nqa, Hubeny:2007xt, Dong:2016hjy,Baggioli:2021xuv}.

Among many different measures of the entanglement, the most commonly used entanglement entropy (EE) can only be used to measure the entanglement between subsystems in pure states. For mixed states, it is necessary to use other kinds of measures, such as entanglement of formation, non-negative, and entanglement of purification \cite{Horodecki:2009review, vidal:2002}.

The holographic duality of EE has been proposed as the holographic entanglement entropy (HEE). HEE is related to the minimum surface in bulk corresponding to the EE in the boundary \cite{Ryu&Takayanagi:2006}. HEE has been studied as the diagnose of quantum and thermodynamic phase transition \cite{Nishioka:2006gr, Klebanov:2007ws, Pakman:2008ui, Zhang:2016rcm, Zeng:2016fsb, Ling:2015dma, Ling:2016wyr, Ling:2016dck, Kuang:2014kha, Guo:2019vni, Mahapatra:2019uql}.  Subsequently, many other holographic quantum measurements have been proposed for mixed states systems. For example, holographic mutual information (MI) is related to the total correlation of quantum systems \cite{hayden2013holographic}. Entanglement of purification and reflect entropy are associated with the minimum area of cross-sections in the entanglement wedge (EWCS) \cite{Takayanagi:2017knl, Chu:2019etd}. EWCS is a novel tool to study the quantum correlation in mixed states systems \cite{Li:2021rff,Liu:2021rks}. Moreover, the quantum butterfly effect is dynamic quantum information properties of a quantum system and related to the geometry of black holes in dual spacetime \cite{Blake:2016wvh, Blake:2016sud, Ling:2016ibq, Ling:2016wuy}.

In general relativity, the no-hair theorem claims that black holes can be determined by mass, charged, and angular-momentum \cite{Israel:1967wq, Carter:1971zc, Ruffini:1971bza}. However, the no-hair theorem is evaded in many new black hole solutions, such as Yang-Mills, Skyrme field, and conformally-coupled scalar field \cite{Volkov:1989fi, Bizon:1990sr, Greene:1992fw, Luckock:1986tr, Bekenstein:1974sf}. Subsequently, a novel no-hair theorem is proposed for the non-minimally coupled scalar field \cite{Bekenstein:1995un}. Moreover, holographic superconductivity is one of the most well-known gravity systems with scalar hair \cite{Hartnoll:2008kx, Cai:2015cya}.

In this paper, we study the mixed state entanglement properties of four-dimension asymptotically AdS black holes with scalar hair \cite{Gonzalez:2013}. The scalar hair of the black holes is minimally coupled to the curvature with a novel potential that allows a simple analytical background solution. The mixed state entanglement properties of this model can help understand the more comprehensive scenario of the role the scalar hair played in determining the mixed entanglement properties.

The paper is organized as followings: we introduce the black holes with scalar hair in \ref{sec: background}, then the concepts, techniques of computation and results of the HEE in \ref{sec: HEE}, the MI (\ref{sec: MI}), and the EWCS (\ref{sec: EWCS}).

\section{Four-Dimensional Asymptotically AdS Black Holes with Scalar Hair}
\label{sec: background}

The action of four-dimensional asymptotically AdS black hole with a scalar hair is \cite{Gonzalez:2013},
\begin{equation}\label{eq:actionn2}
  S=\int d^{4} x \sqrt{-g}\left(\frac{R-2\Lambda}{2 \kappa} -\frac{1}{2} g^{\mu \nu} \nabla_{\mu} \phi \nabla_{\nu} \phi-V(\phi)\right),
\end{equation}
where $\Lambda \equiv -3l^2$ and $\kappa \equiv 8\pi G_N$, and $G_N$ is the Newton constant. $l$ is the length of the AdS space, and $\mu$ is a parameter proportional to the mass. The potential takes the form,
  \begin{equation}\label{eq:vform}
    V ( \phi ) = - \frac { 3 } { 4 \pi G_N l ^ { 2 } } \sinh ^ { 2 } \sqrt { \frac { 4 \pi G_N } { 3 } } \phi.
  \end{equation}
\begin{figure}
  \centering
  \includegraphics[width=0.6\textwidth]{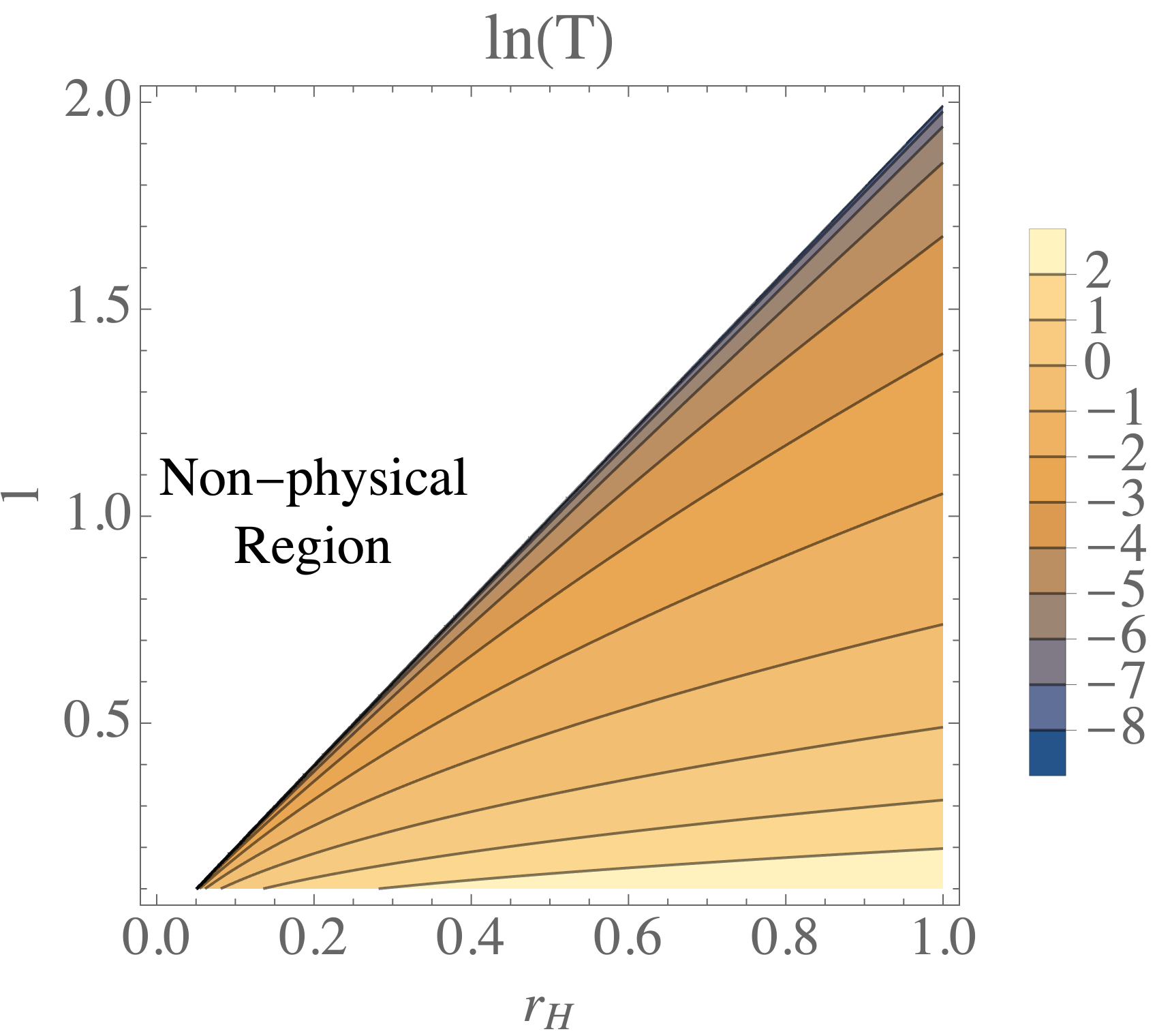}
  \caption{Hawking temperature with parameters $l$ and $r_H$, and the non-physical region is where the temperature is lower than zero.}
  \label{fig: domains_lT}
\end{figure}
As a result, the Einstein equation and energy momentum tensor $T_{\mu \nu}^{(\Phi)}$ are:
\begin{equation}\label{eq:matrixpoly}
  \begin{aligned}
    R_{\mu \nu}-\frac{1}{2} g_{\mu \nu} R                                  & =\kappa T_{\mu \nu}^{(\phi)} ,                                                   \\
    T_{\mu \nu}^{(\phi)} =\nabla_\mu\phi\nabla_{\nu}\phi-g_{\mu \nu} & \left[\frac{1}{2}g^{\rho\sigma}\nabla_\rho \phi\nabla_\sigma\phi+V(\phi)\right].
  \end{aligned}
\end{equation}
The above system admits an analytical solution with an asymptotic AdS, 
\begin{equation}\label{eq:metric_solution}
  d s^{2}=B\left(r\right)\left(-F\left(r\right) d t^{2}+\frac{1}{F\left(r\right)} d r^2+r^2 d \sigma^{2}\right),
\end{equation}
where
\begin{equation}\label{BF_function}
  \begin{aligned}
    B\left(r\right) & =\frac{r\left(r+2 G_{N} \mu\right)}{\left(r+G_{N} \mu\right)^{2}},                                            \\
    F\left(r\right) & =\frac{r^{2}}{l^{2}}-\left(1+\frac{G_{N} \mu}{r}\right)^{2},                                                  \\
    \phi(r)         & = \sqrt { \frac { 3 } { 4 \pi G_N } } \operatorname { arctanh } \left(\frac { G \mu } { r + G_N \mu }\right),
  \end{aligned}
\end{equation}
On the horizon $F(r_H)=0$ can be used to further simplify the parameter space of the system. By solving the equation $F(r)\mid_{r=r_{H}}=0$, one finds that parameter ${\mu=\frac{-l \ r_H+r_H ^2}{l \ G_N}}$.
The Hawking temperature reads $T=\frac{1}{2\pi l}(\frac{2r_H}{l}-1)$.
The system parameters can now be represented by $l$ and $T$. We only consider systems with positive Hawking temperature and positive horizon radius. This imposes a restriction on the parameter space $r_H \geqslant l/2$, which is also depicted in the allowed region of the parameter space (Fig. \ref{fig: domains_lT}) .

\section{The Holographic entanglement entropy}
\label{sec: HEE}

\begin{figure}
  \centering
  \includegraphics[width =0.4\textwidth]{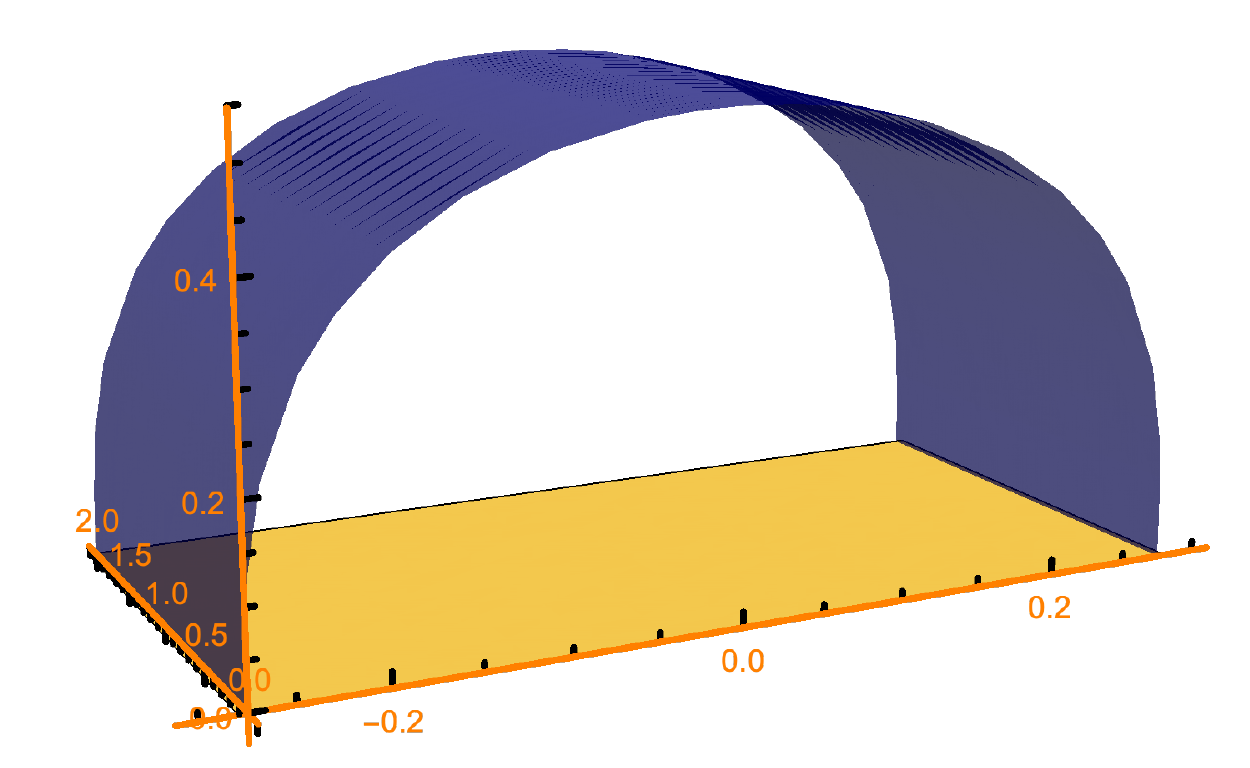}\quad
  \includegraphics[width =0.5\textwidth]{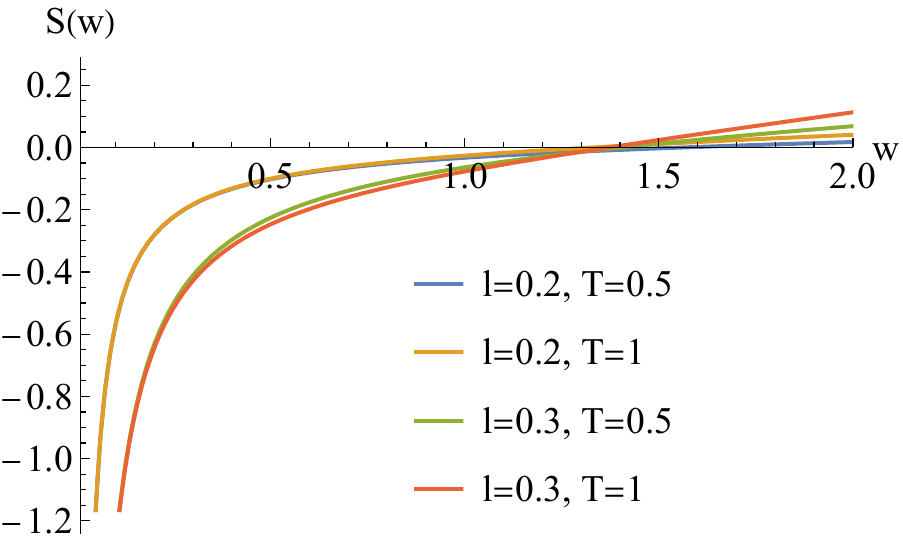}
  \caption{The left plot: the schematic plot of minimum surface for given width ($w$) on $x$ axis.
    The right plot: HEE ($S$) with respect to width ($w$) when parameter $l$ and temperature $T$ are fixed.}
  \label{HEE_demo}
\end{figure}

The EE is defined as,
\begin{equation}\label{EE_def}
  S_{A} (|\psi\rangle) = - \text{Tr}\left[ \rho_{A} \log \rho_{A} \right],
\end{equation}
where the reduced density matrix $\rho_{A} = \text{Tr}_{B} \left(|\psi\rangle\langle\psi|\right)$. In pure state one can find that $S_A = S_B$.
The holographic dual of EE has been proposed in \cite{Ryu&Takayanagi:2006}, and the definition of HEE in pure states reads,
\begin{equation}\label{HEE_def}
  S = \frac{\text{Area}(\Sigma)}{4 \ G_N},
\end{equation}
where $\Sigma$ is the minimum surface in the bulk (see the left plot in Fig. \ref{HEE_demo}),
which is the four-dimensional asymptotically AdS black hole \cite{Gonzalez:2013} in this paper.

In principle, solving the minimum surface can be very complicated. Because it may involve the partial differential equations. However, considering the subregions as infinite strips can turn this problem into the ordinary differential equations, which greatly simplifies the calculation. Here, we let the infinite strip extends to infinity in the $y$ direction and has a finite width in the $x$ direction. Now, the area of the minimum surface is given by,
\begin{equation}\label{Msurface_def}
  \text{Area}(\Sigma)=\int_{\Sigma} \mathcal{L} \ dx dy=\sigma_y \int_{\Sigma} \mathcal{L} \ dx=\sigma_y \int_{\Sigma} \ \frac{\mathcal{L}}{z'(x)} \ dz,
\end{equation}
where $\sigma_y$ is the constant infinite length along $y$-direction. We omit this common infinite constant $\sigma_y$ for simplicity, and treat
\begin{equation}\label{eq:heedef}
  S = \int_0^{z_*} \left(\frac{\mathcal{L}}{z'(x)} -\frac{l^2 (2 \pi  l T+1)}{2 z^2} \right)dz -\frac{l^2 (2 \pi  l T+1)}{2 z_*},
\end{equation}
as the HEE, where the second term in the integral is introduced to subtract out the divergence, and the last term is to compensate the integral. In addition, due to the existence of asymptotic AdS, the integral of Eq. \eqref{Msurface_def}  diverges. Therefore, we adopt the regulation by subtracting a common divergence term to obtain a finite HEE. The right plot in Fig. \ref{HEE_demo} is the relation between $S$ and width $w$, where
\begin{equation}\label{width_def}
  w=\int_{\Sigma} \ dx =\int_{\Sigma} \ \frac{1}{z'(x)} \ dz.
\end{equation}
The entanglement entropy increases with the increase of the width, which is obviously in line with the physical expectation. Because larger sub-regions mean more degrees of freedom to participate in entanglement, resulting in larger entanglement entropy.

\begin{figure}
  \includegraphics[width =0.49\textwidth]{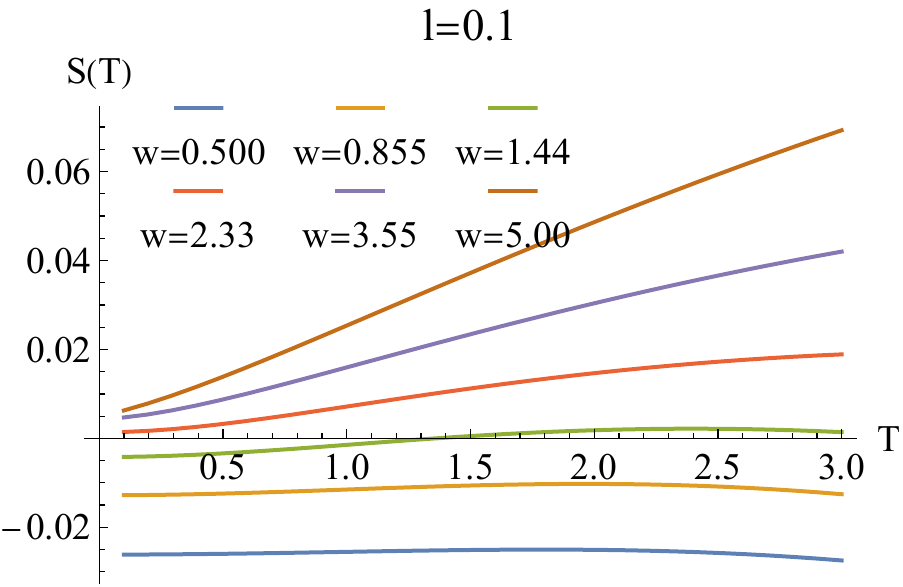}
  \includegraphics[width =0.49\textwidth]{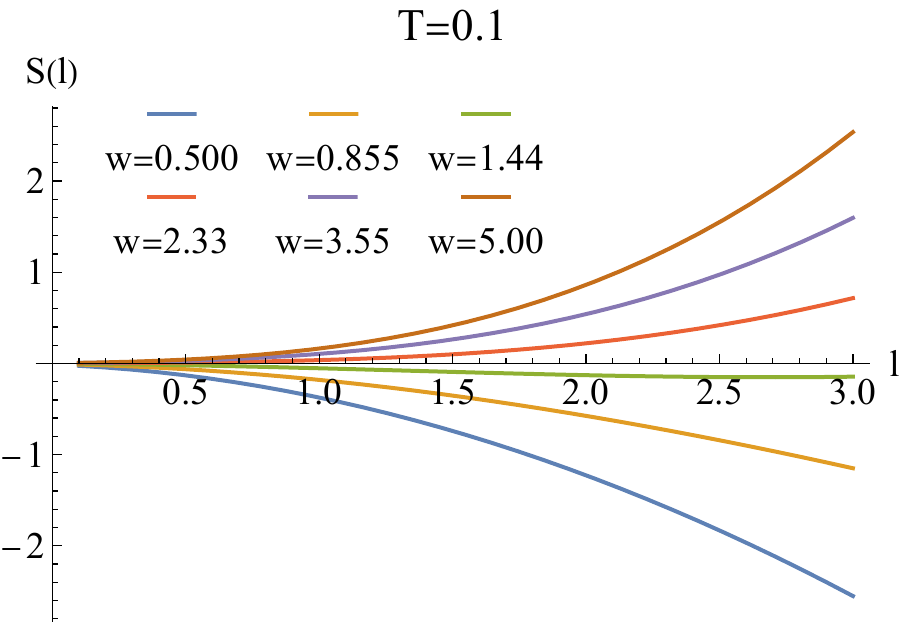}
  \caption{HEE vs $T$ (left plot) and HEE vs $l$ (right plot).}
  \label{HEE_even}
\end{figure}

With the data in Fig. \ref{HEE_demo}, one can study the relationship between the HEE and $T$, as well as the relationship between the $S$ and $l$, at fixed values of width. We show $S$ vs $T$ and $S$ vs $l$ in Fig. \ref{HEE_even}. As can be seen from the Fig. \ref{HEE_even}, the behavior of $S$ with $T$ is similar to that of $S$ with $l$. When the width is small, $S$ decreases monotonically with $T$ and $l$; when the width gradually increases, the behavior of $S$ with $T$ and $l$ becomes non-monotonic; finally, when the width is large enough, $S$ increases monotonically with $T$ and $l$. It is worth noting that when the width is large, $S$ linearly increases with $T$, while $S$ increases non-linearly with $l$. Next, we understand these phenomena from both physical and analytical perspectives.

Firstly, the relationship between HEE and temperature is opposite to that of RN-AdS at small width \cite{Liu:2019qje}. Here, we also use a similar analytical method to understand this.
When the width is small, the minimum surfaces exist in small $z$ regions. Therefore, we expand the integral with $z$ when calculating the minimum surfaces.
\begin{equation}\label{eq:heeexp1}
  S = \int_0^{z_*} \left(\frac{-16 \pi ^4 l^6 T^4+8 \pi ^2 l^4 T^2+7 l^2}{16 \pi  l T+8}+O\left(z^1\right)\right)dz-\frac{l^2 (2 \pi  l T+1)}{2 z_*}.
\end{equation}
According to Eq. \eqref{eq:heedef}, the partial derivative of $S$ to $T$ and $l$ are,
\begin{equation}\label{eq:heed1}
  \begin{aligned}
    \partial_T S & = -\frac{\pi  l^3}{z_*} + \int_0^{z_*}  \left(\frac{1}{4} \pi  l^3 \left(4 \pi  l T (1-3 \pi  l T)-\frac{8}{(2 \pi  l T+1)^2}+1\right)\right)dz,                   \\
    \partial_l S & = -\frac{l (3 \pi  l T+1)}{z_*} + \int_0^{z_*} \left(\frac{l (\pi  l T (8 \pi  l T (\pi  l T (3-2 \pi  l T (5 \pi  l T+3))+2)+7)+7)}{4 (2 \pi  l T+1)^2}\right)dz.
  \end{aligned}
\end{equation}
Because $z$ is small, $z_*$ is also small. Thus the above equation can be approximated as,
\begin{equation}\label{eq:heed2}
  \begin{aligned}
    \partial_T S & \simeq -\frac{\pi  l^3}{z_*} <0,         \\
    \partial_l S & \simeq -\frac{l (3 \pi  l T+1)}{z_*} <0.
  \end{aligned}
\end{equation}
This explains the phenomenon that $S$ decreases monotonically with $T$ and $l$ when the width is small.

Next, we explain the increasing behavior of $S$ with $T$ and $l$ at large width. When the width is large enough, $S$ will approach,
\begin{equation}\label{eq:Ssw}
  S \simeq sw,
\end{equation}
where $s$ is the entropy density,
\begin{equation}\label{eq:entropydensity}
  s = 2 \pi  l^3 T.
\end{equation}
Therefore, the behavior of $S$ at large width is mainly determined by thermal entropy. Obviously, $S$ is proportional to $T$ and $l$, which explains the increasing behavior of $S$ with $T$ and $l$ at large width. Furthermore, Eq. \eqref{eq:entropydensity} shows that $S$ is linear to $T$, while $S$ is linear to $l^3$. Therefore, the Eq. \eqref{eq:entropydensity} also explains the linear growth of $S$ with $T$ and the nonlinear growth of $S$ with $l$ at large width.

\section{The holographic mutual information}
\label{sec: MI}

\begin{figure}
  \includegraphics[width=0.49\textwidth]{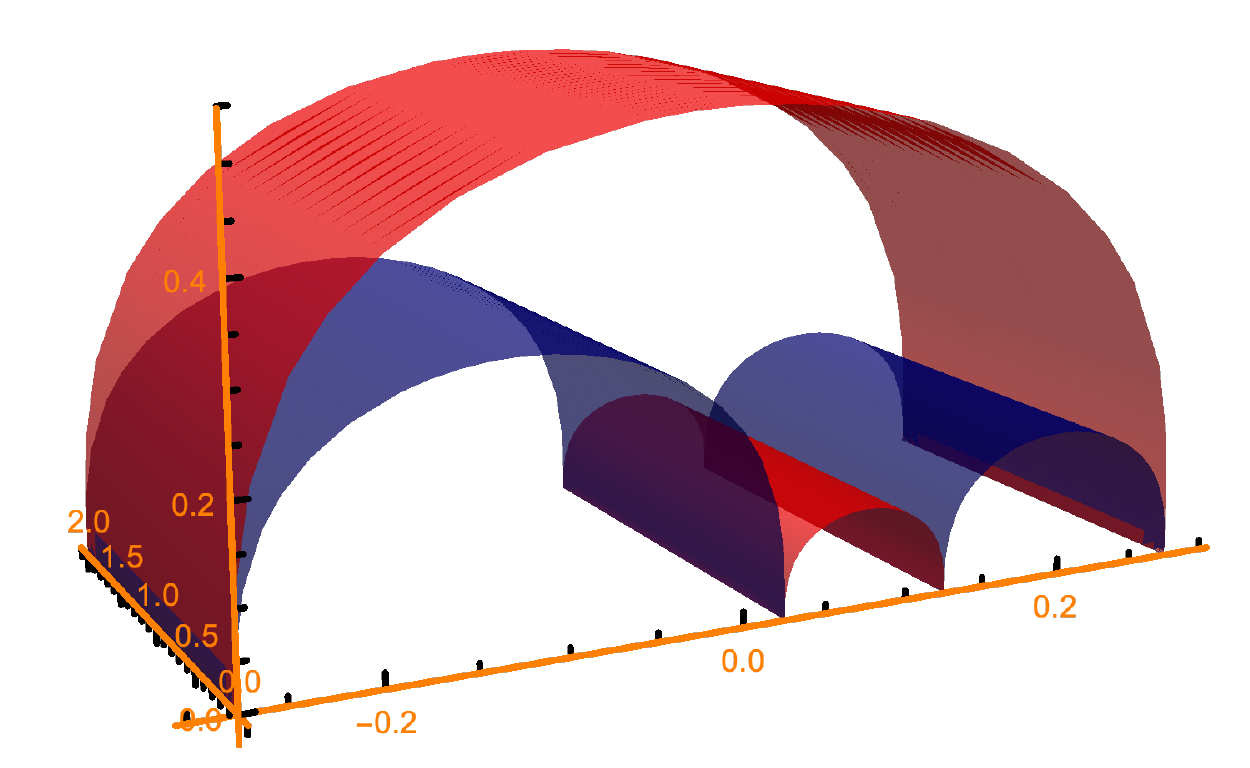}
  \includegraphics[width=0.49\textwidth]{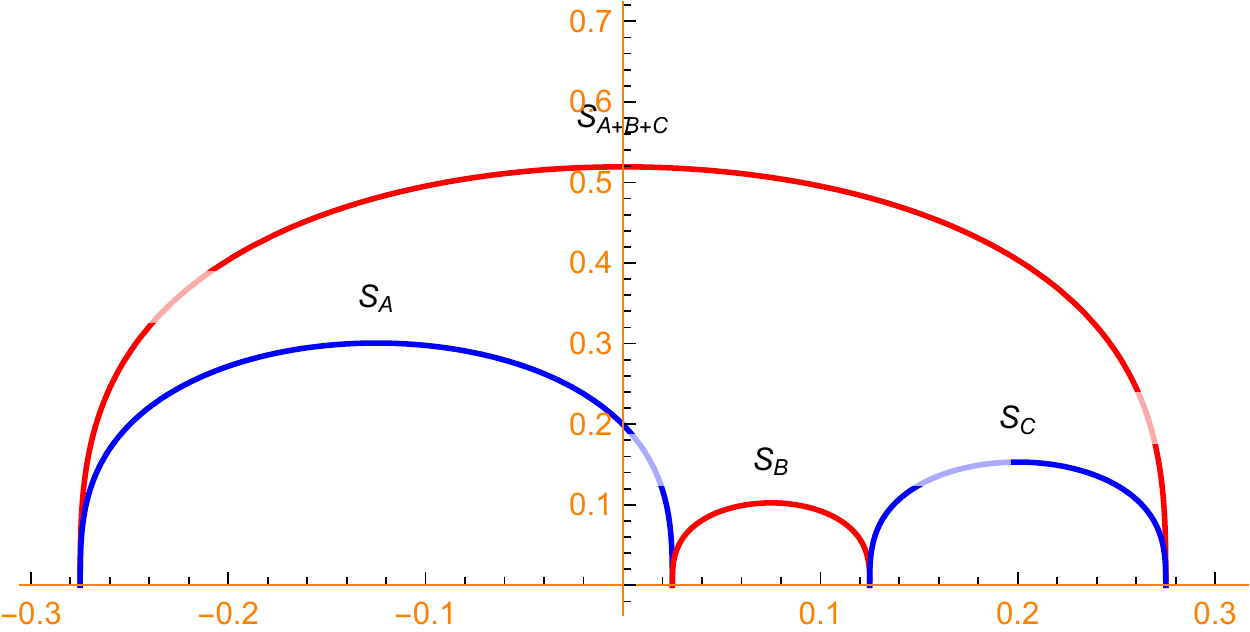}
  \caption{The demonstration of MI in 3D and 2D, the colors are identical.}
  \label{fig: MI_demo}
\end{figure}

The mutual information (MI) is defined below \cite{hayden2013holographic},
\begin{equation}\label{mi:def}
  I\left(A,C\right) := S\left(A\right) + S\left(C\right) - S\left(A\cup C\right),
\end{equation}
where $A$ and $C$ are two disjoint subsystems. We show this in Fig. \ref{fig: MI_demo}, in which the left plot is the 3D demonstration and the right plot the 2D section plot with the help of the infinite strip configuration. By definition, the value of MI is directly a linear combination of HEE,
\begin{equation}\label{mi:def_holo}
  I\left(A,C\right)=S_A+S_C-S_{A\cup B\cup C}-S_B.
\end{equation}
The configuration of the MI can be denoted as $(a,b,c)$, the widths of $A,\,B$ and $C$. The MI can partially subtract out the thermal contribution from the HEE, and qualifies a better diagnose of mixed state entanglement measure than HEE \cite{Fischler:2012}. Next, we discuss the relationship between MI and temperature.

\begin{figure}
  \centering
  \includegraphics[width=0.45\textwidth]{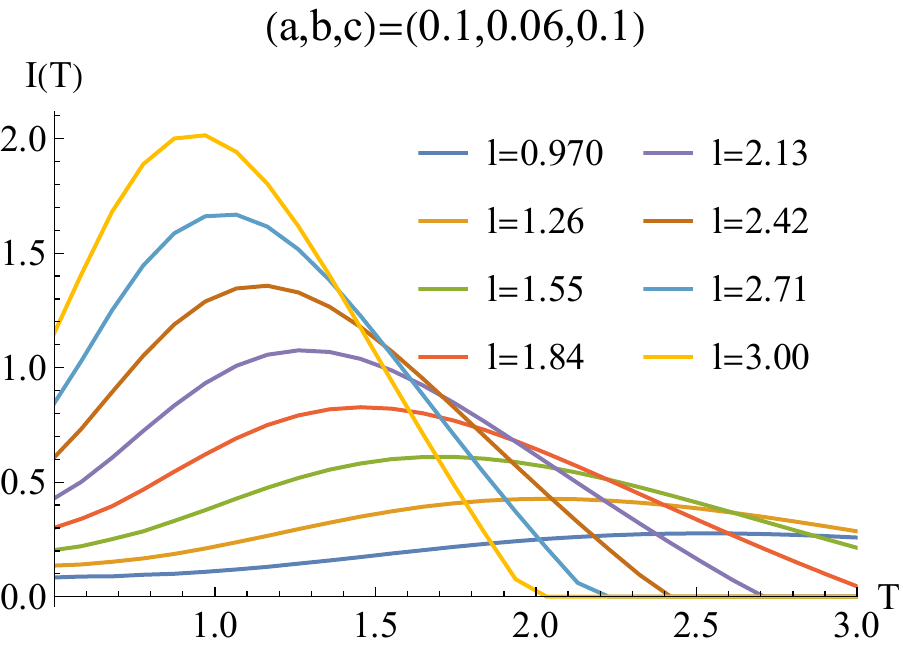}
  \includegraphics[width=0.45\textwidth]{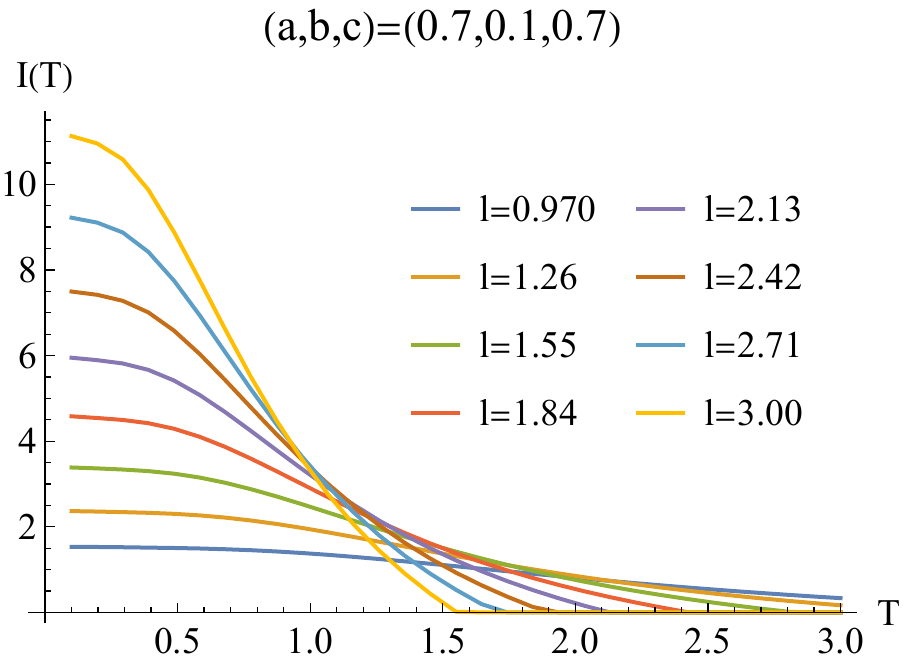}
  \caption{MI vs $T$ for small configuration (left plot) and large configuration (right plot). Different curves in each plot correspond to different values of $l$ represented by the plot legends.}
  \label{fig:ivst}
\end{figure}

In Fig. \ref{fig:ivst}, we show the relationship between MI and temperature for small configuration (left plot) and large configuration (right plot).
First, when the configuration is small (see the left plot of Fig. \ref{fig:ivst}), the MI monotonically increases with temperature when $l$ and $T$ are small. This is a manifestation of the monotonic decrease of HEE with temperature at small width. Because when the width is small, the minimum surface will reside near the AdS boundary. Therefore, MI exhibits the opposite behavior as the largest minimum surface since it can more comprehensively capture the deviation from the AdS. That is, MI behavior is opposite to that of the $S_{A\cup B\cup C}$. As $l$ increases, MI can gradually show non-monotonic behavior with $T$. When $l$ and $T$ further increase, MI shows non-monotonic behavior with $T$. This stems from the non-monotonic behavior of HEE with $T$ and $l$ when $T$ and $l$ are not small (see Fig. \ref{HEE_even}). When the temperature is high enough, the MI will gradually vanish. This is because the high temperature will destroy the entanglement between different sub-regions. From the perspective of statistical physics, when the temperature is high, the reduced density matrix of subsystems will be closer to a thermal density matrix, resulting in disentangling subregions.

When the configuration is large (the right plot of Fig. \ref{fig:ivst}), however, the MI decreases monotonically with the temperature and vanishes when the temperature is large. This is because, for large configurations, the minimum surfaces corresponding to $S_A$, $S_C$ and $S_{A\cup B\cup C}$ are close to the horizon of the black brane. At this time, the minimum surface of $S_B$ cannot be too large, otherwise, MI will vanish. Therefore, from Eq. \eqref{mi:def} we can see that, for large configurations, MI is determined by $S_ B$, which shows a monotonic decreasing behavior.

\begin{figure}
  \centering
  \includegraphics[width=0.45\textwidth]{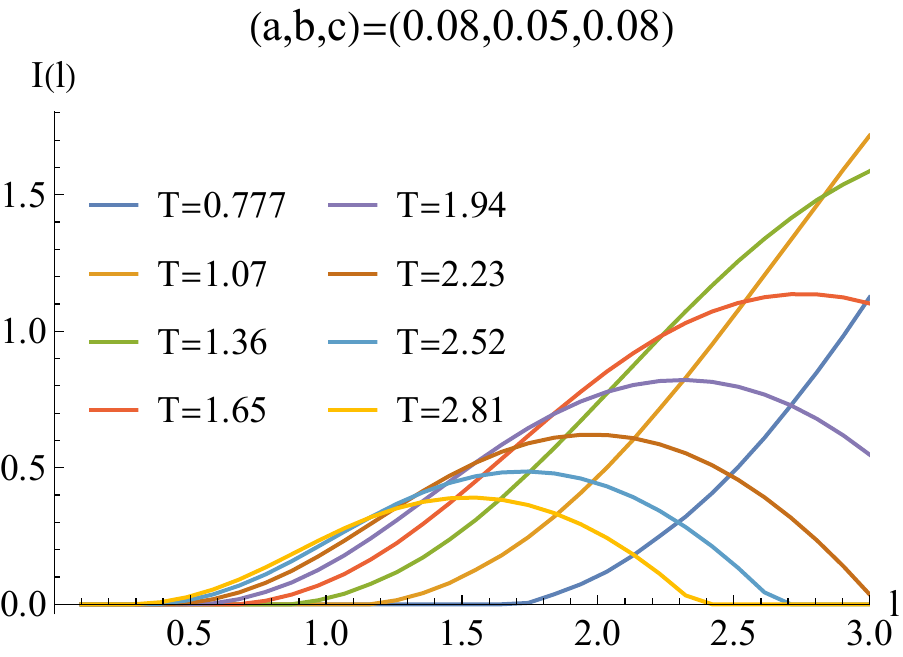}
  \includegraphics[width=0.45\textwidth]{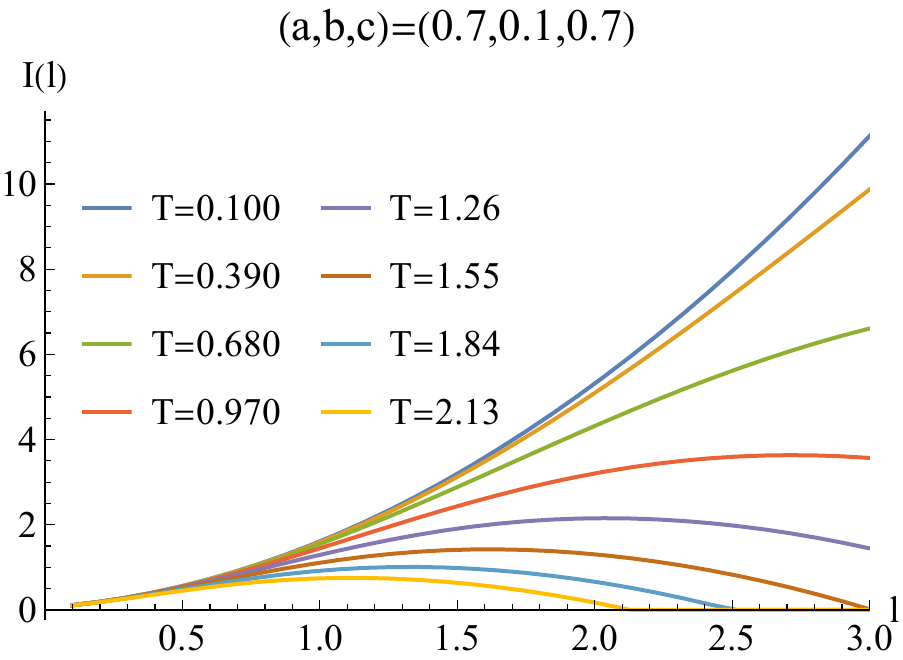}
  \caption{The MI vs $l$ for small configuration (left plot) and large configuration (right plot). Different curves in each plot correspond to different values of $T$ represented by the plot legends.}
  \label{fig:ivsl}
\end{figure}

Next, we study the relationship between MI and $l$. We show the results in Fig. \ref{fig:ivsl},
where the left and right plot are the MI vs $l$ for small configurations and large configurations. First, the MI vanishes when $l$ is small. When $l$ increases gradually, MI increases with the increase of $l$. At this time, this increasing behavior lasts for a large range of $l$ when the temperature is relatively small. However, when the temperature further increases, MI becomes non-monotonic with $l$ and vanishes for large values of $l$. Increasing the temperature or $l$ will enhance the horizon radius $r_H$, hence making the minimum surface closer to the horizon. In this situation, the MI is determined by $-S_{A\cup B \cup C}$, where a decreasing behavior with $l$ will be anticipated until MI vanishes.

\section{The Entanglement Wedge Minimum Cross-Section}
\label{sec: EWCS}

\begin{figure}
  \includegraphics[width=0.49\textwidth]{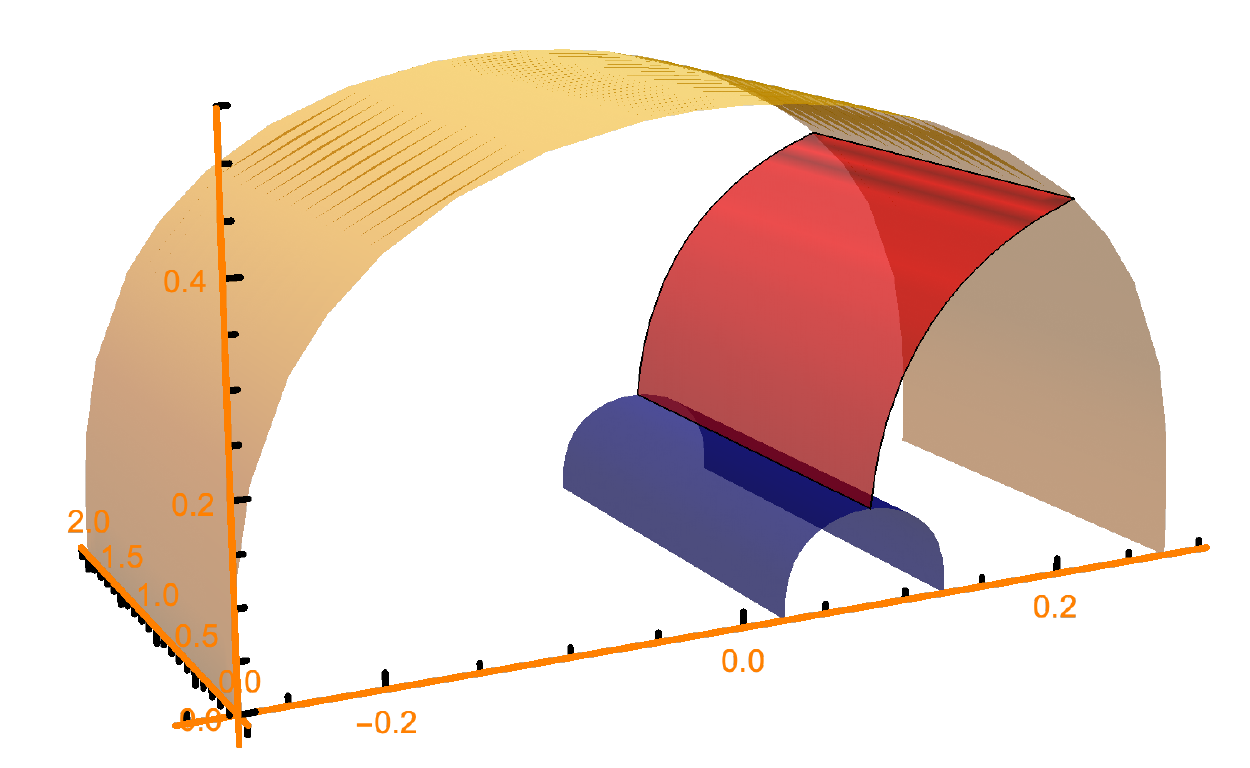}
  \includegraphics[width=0.49\textwidth]{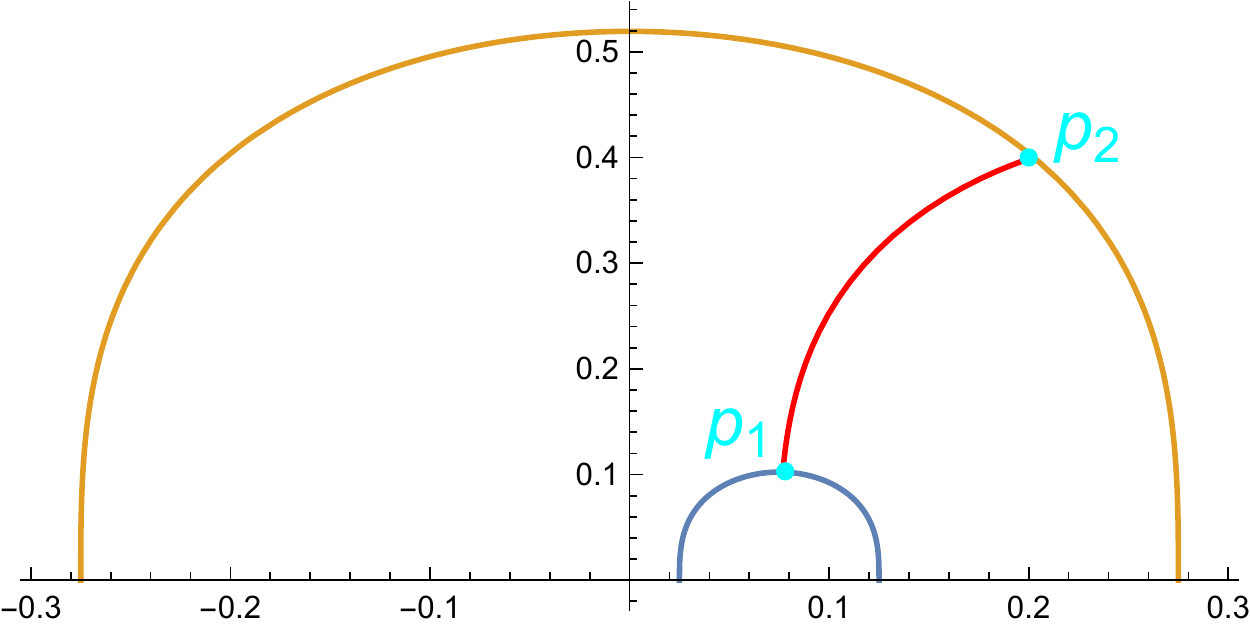}
  \caption{The demonstration of EWCS in $3D$ and $2D$, the colors are identical.}
  \label{fig: EoP_demo}
\end{figure}

The entanglement of the mixed state can be measured by $E_W(\rho_{AB})$, which is proportional to the minimum cross-section $\Sigma_{AB}$ in the entanglement wedge \cite{Takayanagi:2017knl},
\begin{equation}\label{equ: ewcsdef}
  E_W(\rho_{AB})=\min_{\Sigma_{AB}}\left(\frac{\text{Area}(\Sigma_{AB})}{4G_N}\right),
\end{equation}
called the entanglement wedge cross-section (EWCS). EWCS has very different geometric properties compared to HEE and MI, as is shown in Fig. \ref{fig: EoP_demo}.

To find out the minimum cross-section, we implement the Newton-Raphson method and boundary condition that the red surface (curve) are perpendicular to the boundaries, the yellow and the blue surfaces (curves) in Fig. \ref{fig: EoP_demo}. After a coordinate transfomation, making $\{t,z,x,y\}\rightarrow\{t,z,\theta,y\}$ with $\theta = \arctan z/x$, we have
\begin{equation}\label{equ: BCs}
  \left\{\begin{array}{l}
    Q_1 \equiv\left.g_{a b}\left(\frac{\partial}{\partial z}\right)^{a}\left(\frac{\partial}{\partial \theta_{1}}\right)^{b}\right|_{P_{1}}=0 \\
    Q_2 \equiv\left.g_{a b}\left(\frac{\partial}{\partial z}\right)^{a}\left(\frac{\partial}{\partial \theta_{2}}\right)^{b}\right|_{P_{2}}=0,
  \end{array}\right.
\end{equation}
where $P_1$ and $P_2$ are are points where the minimum cross-section (curve) adjoin two others.

Next, the computation of EWCS is followed by the Newton-Raphson method, which could be written as
\begin{equation}
  \left(\begin{array}{l}
    Q_{1} \\
    Q_{2}
  \end{array}\right)+\left(\begin{array}{ll}
    \partial_{\theta_{1}} Q_{1} & \partial_{\theta_{2}} Q_{1} \\
    \partial_{\theta_{1}} Q_{2} & \partial_{\theta_{2}} Q_{2}
  \end{array}\right)\left(\begin{array}{l}
    \delta \theta_{1} \\
    \delta \theta_{2}
  \end{array}\right)
  =0,
\end{equation}
where $Q_1$ and $Q_2$ are equation (\ref{equ: BCs}) after normalization so as to facilitate computations.
Given a proper first guess of $(\theta_1, \theta_2)$, one can obtain feedbacks $(\delta \theta_1,\delta \theta_2)$ to iterate until the condition \eqref{equ: BCs} is satisfied.

\begin{figure}
  \includegraphics[width=0.49\textwidth]{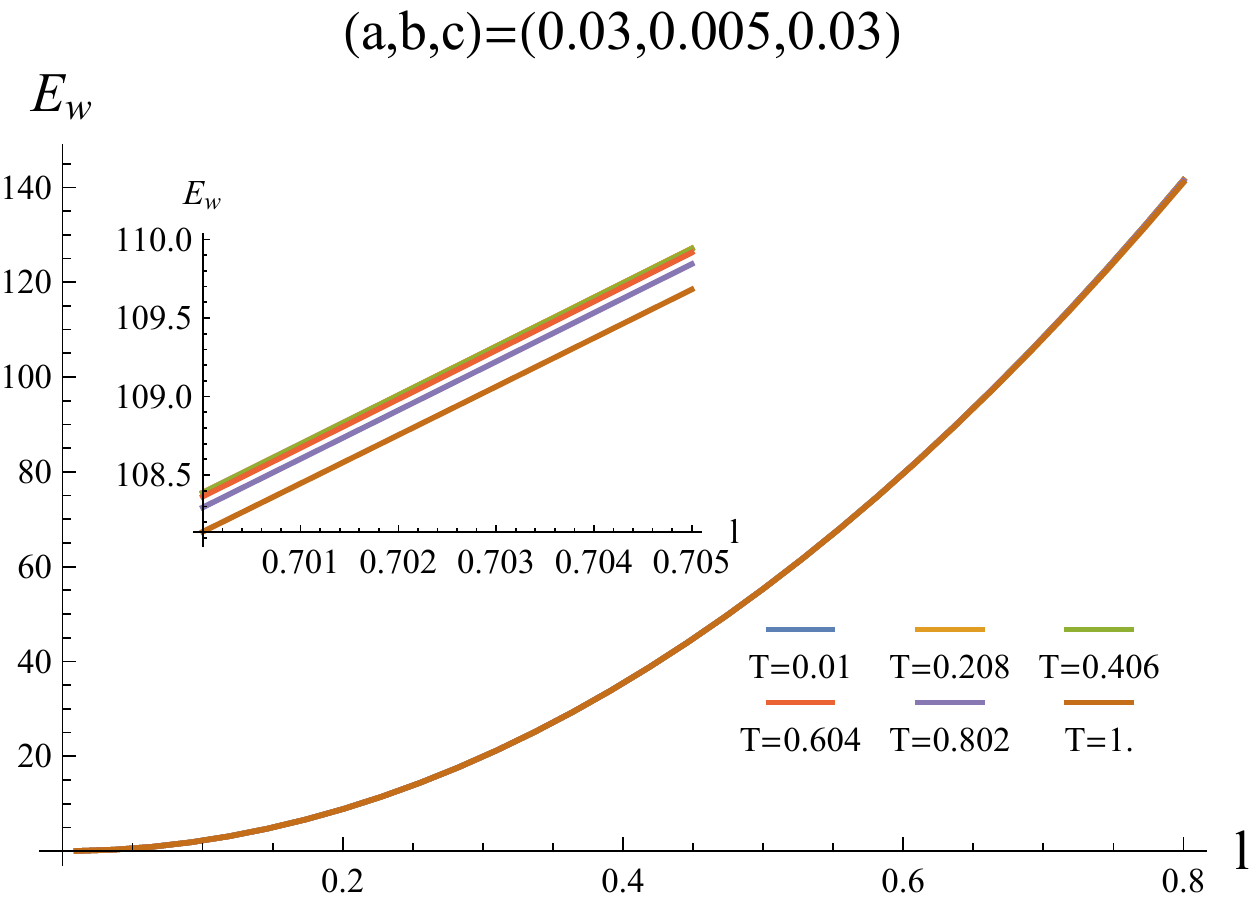}
  \includegraphics[width=0.49\textwidth]{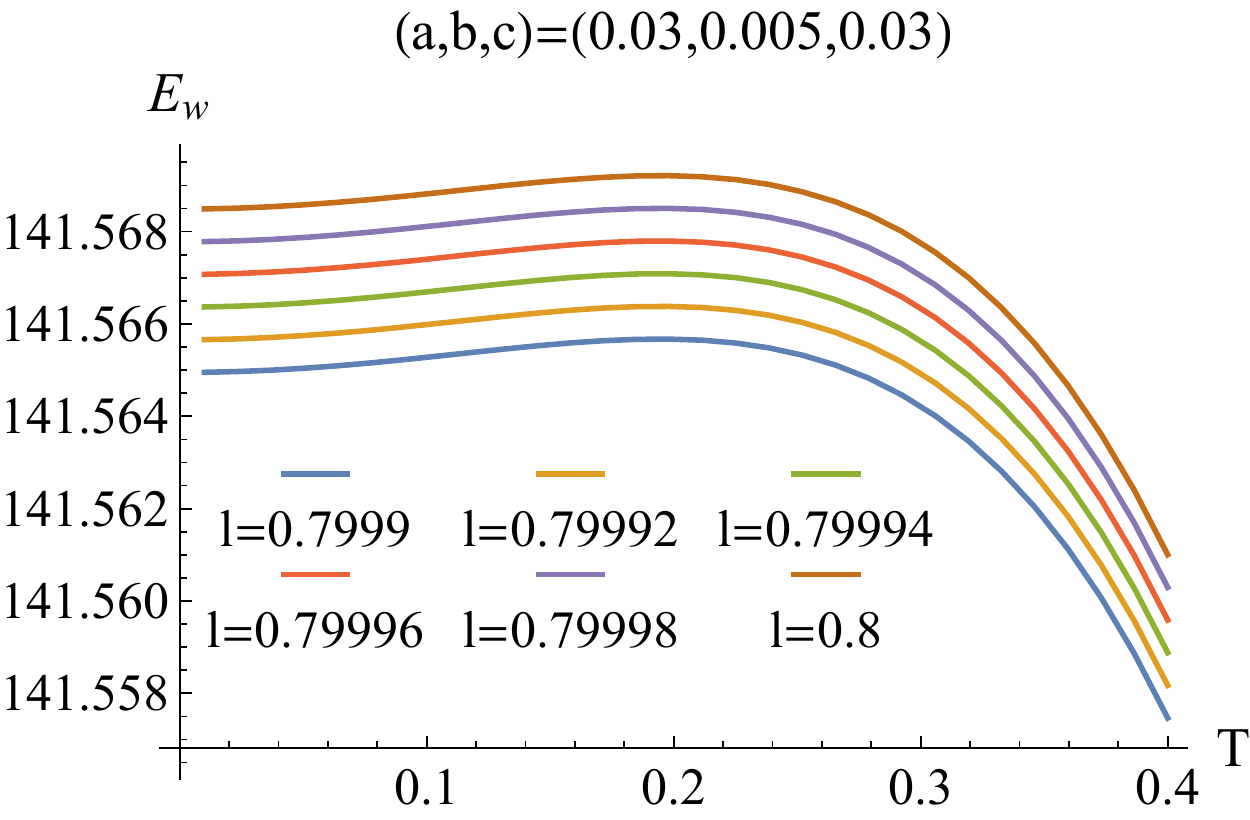}
  \caption{$E_W$ vs $l$ and $T$ respectively. $E_W$ only varies slightly with $T$ for small configurations that the curves(in the left plot) almost overlap with each other. } 
  \label{fig:ssc}
\end{figure}

First, we study the EWCS behavior vs $T$ and $l$ for small configurations. From the Fig. \ref{fig:ssc} we find that EWCS increases with $l$ monotonically, and the $E_W$ of $T$ only shows the monotonicity of increase when the temperature $T$ is low. This means that for small range degrees of freedom, the mixed state entanglement increases with the temperature as well as the AdS radius.

Next, we provide an analytical understanding on the monotonically increasing behavior in the previous paragraph. Under a small variation of $T$ or $l$, we label the variation of $E_W$ as,
\begin{equation}
  \delta E_W=\delta E_W^{(\text{cs})} + \delta E_W^{(\text{m})},
\end{equation}
where the $E_W^{(\text{cs})}$ is the contribution from the change of the cross-section, and the $E_W^{(\text{m})}$ is the contribution from the deformation of the metric. The former one $\delta E_W^{(\text{cs})}$ vanishes following the definition of the minimum cross-section, which means that we only need to variate the integrand expression of the EWCS. The EWCS can be written as,
\begin{equation}\label{Ew}
  \begin{aligned}
    E_{W} & =\int \sqrt{g_{y y}} \sqrt{g_{x x} d x^{2}+g_{z z} d z^{2}} \\
          & = \int \sqrt{\Omega(z) d x^{2}+\Xi(z) d z^{2}},
  \end{aligned}
\end{equation}
where $\Omega(z)\equiv g_{yy} g_{xx}$ and $\Xi(z)\equiv g_{yy} g_{zz}$. Taking the following derivative,
\begin{equation}\label{Ew_mono}
  \begin{aligned}
     & \frac{\partial E_{W}}{\partial T}=\int \frac{\partial}{\partial T} \sqrt{\Omega(z) d x^{2}+\Xi(z) d z^{2}}
    \\
     & \frac{\partial E_{W}}{\partial l}=\int \frac{\partial}{\partial l} \sqrt{\Omega(z) d x^{2}+\Xi(z) d z^{2}},
  \end{aligned}
\end{equation}
the monotonicity can be obtained from the sign of the above two expressions. We can expand $\{ {\frac{\partial \Omega(z)}{\partial T},\frac{\partial \Xi(z)}{\partial T}}\}$ and $\{ {\frac{\partial \Omega(z)}{\partial l},\frac{\partial \Xi(z)}{\partial l}}\}$ near $z=0$ because the configurations are small,
\begin{equation}\label{Ew_monoo}
  \begin{aligned}
     & \frac{\partial \Omega(z)}{\partial T}=\frac{l^{5} \pi(1+2 l \pi T)^{3}}{2 z^{4}}+\frac{1}{O[z]^{2}}
    \\
     & \frac{\partial \Xi(z)}{\partial T}=\frac{l^{5} \pi(1+2 l \pi T)}{z^{4}}+\frac{1}{O[z]^{2}}
    \\
     & \frac{\partial \Omega(z)}{\partial l}=\frac{l^{3}(1+2 l \pi T)^{3}(1+4 l \pi T)}{4 z^{4}}+\frac{1}{O[z]^{2}}
    \\
     & \frac{\partial \Xi(z)}{\partial l}=\frac{l^{3}\left(1+5 l \pi T+6 l^{2} \pi^{2} T^{2}\right)}{z^{4}}+\frac{1}{O[z]^{2}},
  \end{aligned}
\end{equation}
from which we could see that for small configurations, both $ {\frac{\partial E_{W}}{\partial T}}$ and $ {\frac{\partial E_{W}}{\partial l}}$ are positive, which means the integrand of $E_W$ increases monotonically with respect to these two parameters. However, when $T$ is relatively large, $E_W$ can decrease with $T$. This may be explained by the $ {\frac{1}{O[z]^2}}$ term, which can leads to a decreasing $E_W$ with $T$.

\begin{figure}
  \includegraphics[width=0.49\textwidth]{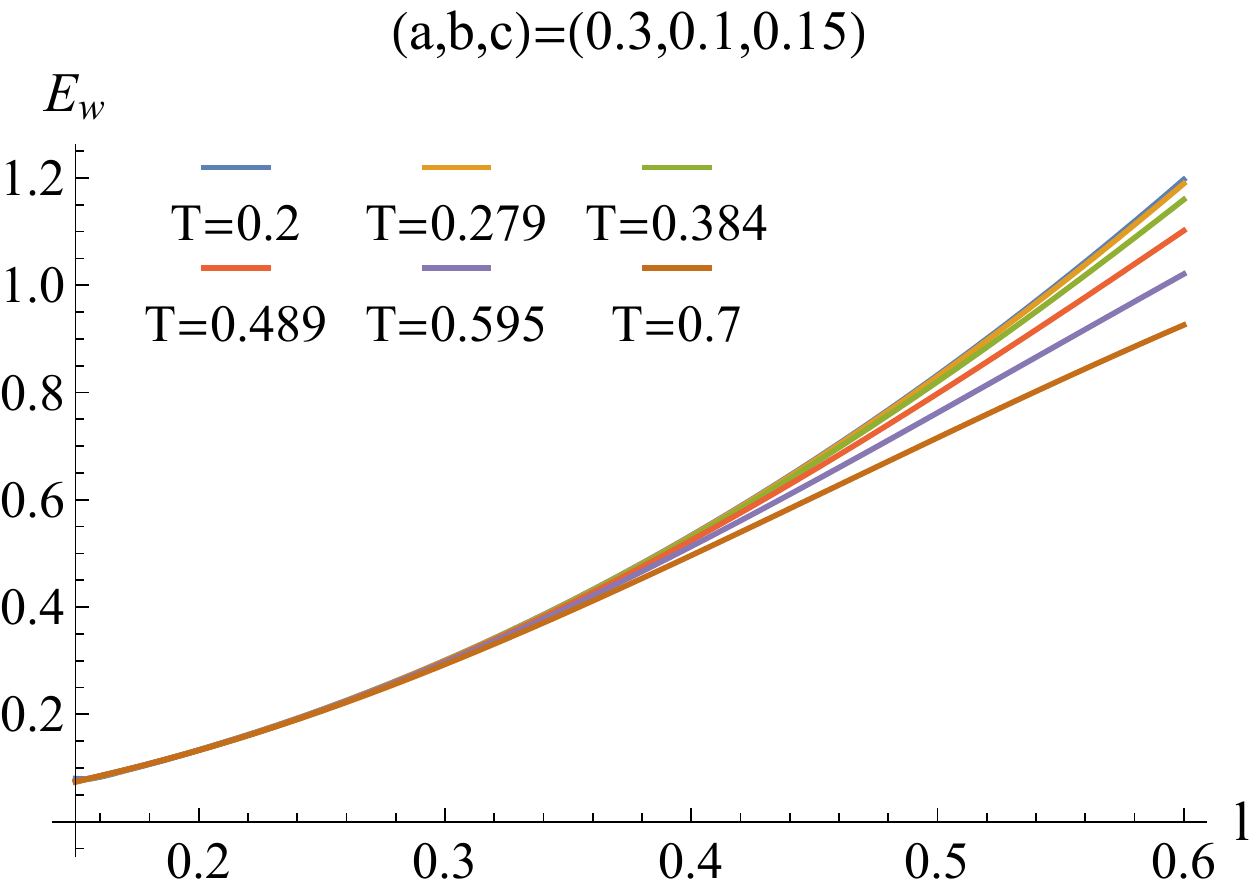}
  \includegraphics[width=0.49\textwidth]{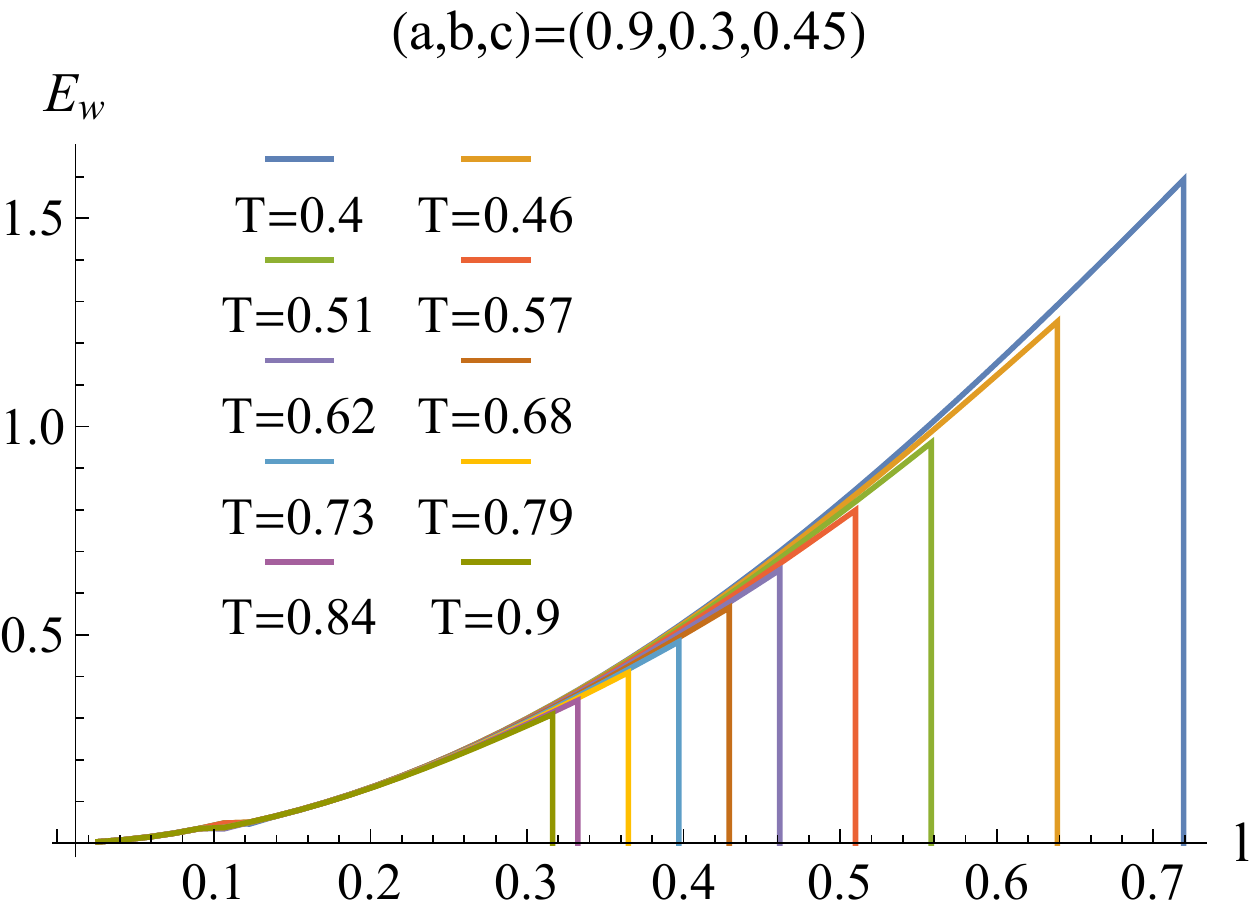}
  \\
  \includegraphics[width=0.49\textwidth]{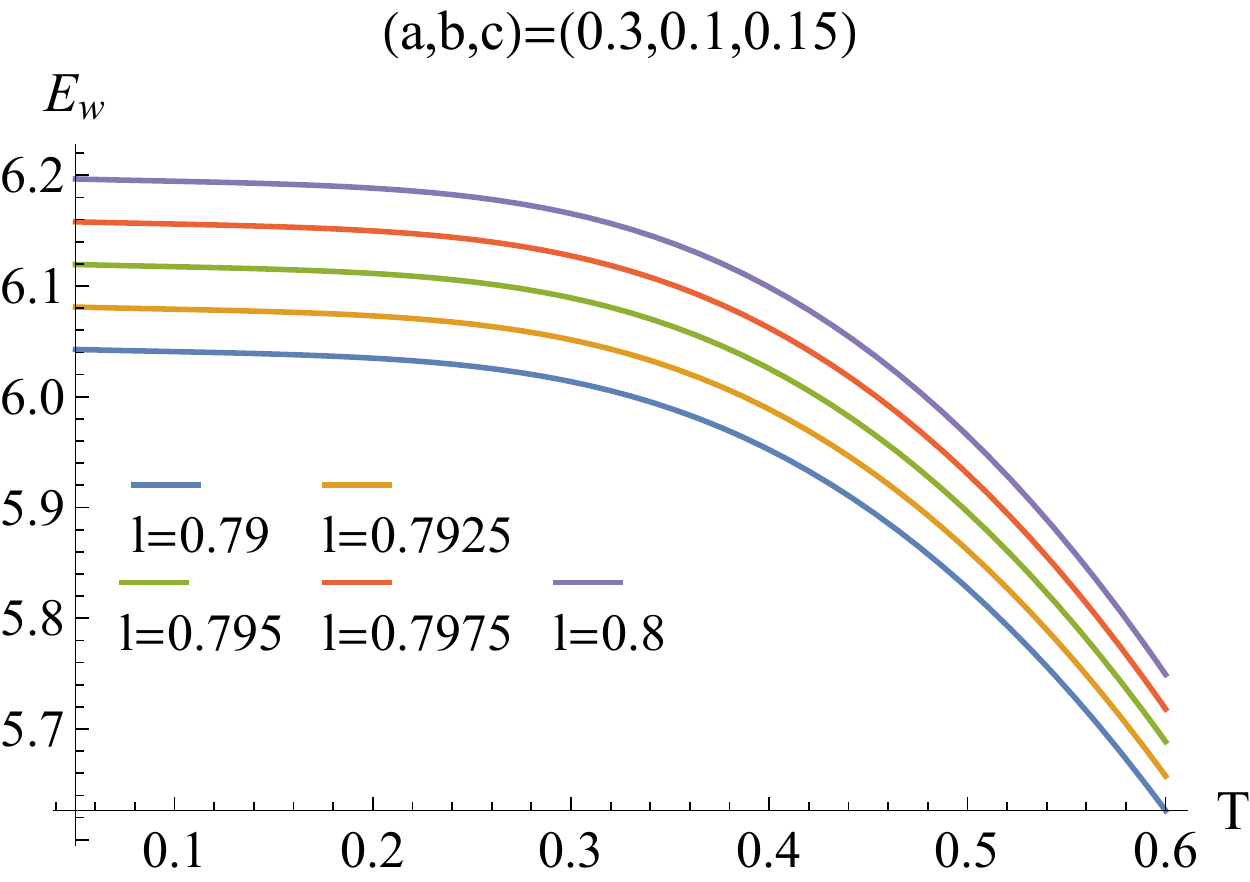}
  \includegraphics[width=0.49\textwidth]{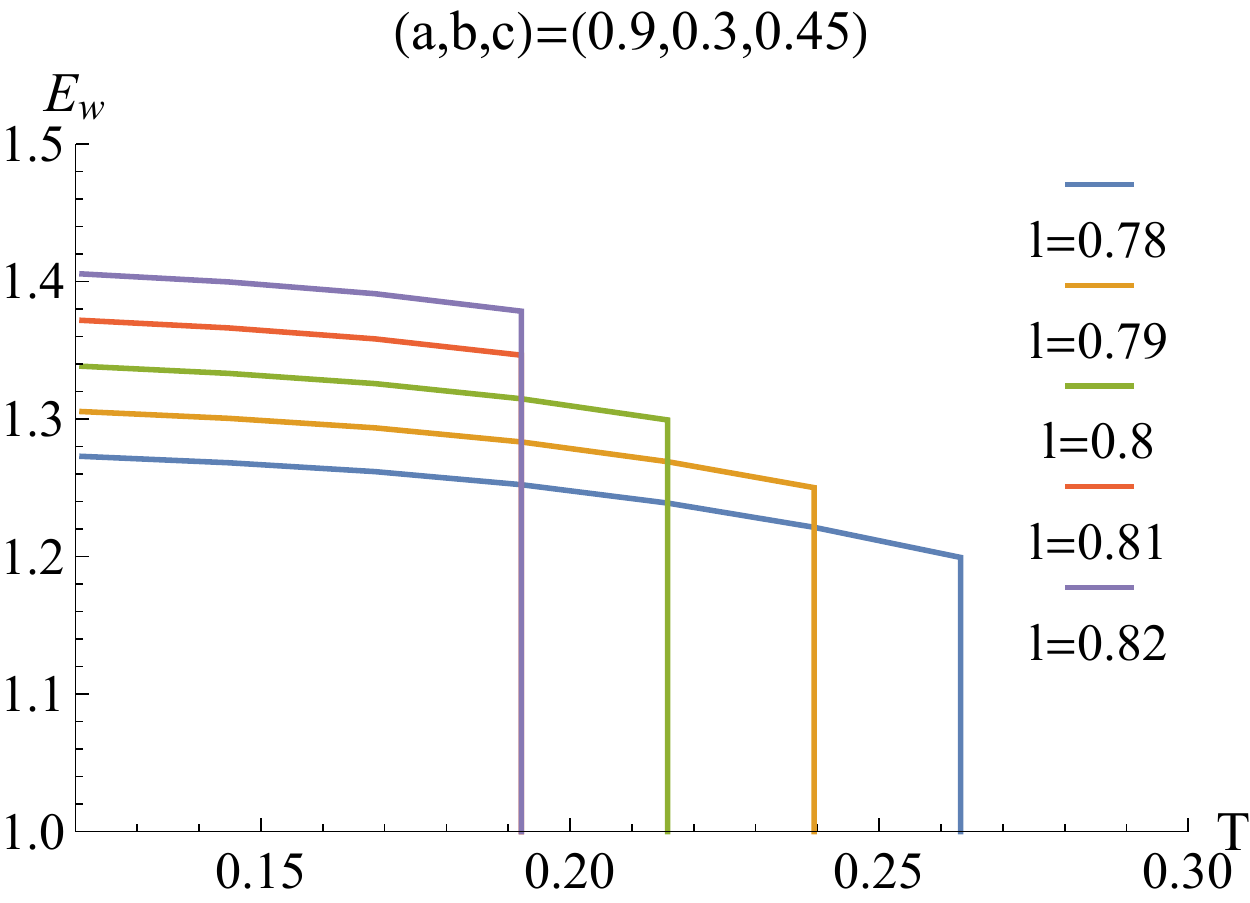}
  \caption{The EWCS of $l$ and $T$. The $E_W$ of larger configuration on the right side will eventually fall to zero because MI vanishes with large parameters.}
  \label{fig:Ew_lT}
\end{figure}

Next, we study the EWCS vs $T$ and $l$ for large enough configurations (Fig. \ref{fig:Ew_lT}). The left column shows the EWCS vs $l$ and $T$ for relatively large configurations where $E_W$ monotonically increases with $l$ but decreases with $T$. This suggests that $E_W$ only exhibits monotonically increasing behavior with $l$, but can exhibit non-monotonic behavior with $T$ since $E_W$ behaves oppositely for small and large configurations. In the right column, we show the $E_W$ vs $l$ and $T$ for even larger configurations, where the monotonicity is the same as the left column, while the $E_W$ can drop to zero as MI vanishes.

\section{Discussion}

We have investigated HEE, MI, and EWCS in infinite strip configurations of four-dimension asymptotically AdS black holes with scalar hair. HEE and MI all show non-monotonic behavior with $T$ and $l$. Moreover, the relationship between HEE and $T$ and $l$, especially the asymptotic linear and cubic relationship, can be understood analytically. Also, the MI behavior can also be understood in an analytical manner because of its dependence on the HEE. The EWCS, however, is distinct from HEE and MI from the definition. We find that the EWCS monotonically increases with $T$ and $l$ for small configurations, but it is non-monotonic with $T$ for intermediate configurations. However, EWCS only exhibits monotonically increasing behavior with $l$. The small configuration case can be understood by analytically working out the derivative with respect to EWCS. The intermediate configuration case, however, can only be captured by numerics so far.

The above phenomena show that the current model with scalar hair shows very different entanglement properties from the AdS-RN model and many other models \cite{Liu:2019qje}. The essence underlying the difference is the novel potential, which leads to the novel dependence of the metric on the temperature and the AdS radius.

Next, we point out several future directions worthy of investigation. It is desirable to compare the current case with the case with superconductivity condensation field. It will offer more insights into the mechanism leading to the distinct behaviors of quantum information revealed in this paper. Another interesting direction is to examine other forms of potential, to see how exactly the potential affects the entanglement properties. This may require numerical treatments for the background solutions since analytical solutions may not be available.

\section*{Acknowledgments}

Peng Liu would like to thank Yun-Ha Zha for her kind encouragement during this work. This work is supported by the Natural Science Foundation of China under Grant No. 11905083.


\begin{thebibliography}{99}
  \bibitem{Osterloh:2002na}
  A. Osterloh, L. Amico, G. Falci, R. Fazio,
  ``Scaling of Entanglement close to a Quantum Phase Transitions''
  Nature {\bf 416}, 608 (2002) [arXiv:0202029 [quant-ph]]

  \bibitem{Amico:2007ag}
  L. Amico, R. Fazio, A. Osterloh and V. Vedral,
  ``Entanglement in many-body systems''
  Rev.Mod.Phys. {\bf 80}, 517 (2008)
  [arXiv:0703044 [quant-ph]]

  \bibitem{Wen:2006topo}
  Levin, Michael, and Xiao-Gang Wen.
  ``Detecting topological order in a ground state wave function''. Physical review letters 96.11 (2006): 110405.

  \bibitem{Kitaev:2006topo}
  Kitaev, Alexei, and John Preskill.
  ``Topological entanglement entropy''. Physical review letters 96.11 (2006): 110404.


  \bibitem{Maldacena:1997}
  Juan M. Maldacena
  ``The Large N Limit of Superconformal Field Theories and Supergravity''
  [arXiv:hep-th/9711200]

  \bibitem{hooft:1993}
  G. 't Hooft
  ``Dimensional Reduction in Quantum Gravity''
  [arXiv:gr-qc/9310026]


  \bibitem{Ryu&Takayanagi:2006}
  Shinsei Ryu and Tadashi Takayanagi,
  ``Holographic Derivation of Entanglement Entropy from AdS/CFT''
  [arxiv:0603001v2]

  \bibitem{Lewkowycz:2013nqa}
  A.~Lewkowycz and J.~Maldacena,
  ``Generalized gravitational entropy,''
  JHEP {\bf 1308}, 090 (2013)
  [arXiv:1304.4926 [hep-th]].

  \bibitem{Hubeny:2007xt}
  V.~E.~Hubeny, M.~Rangamani and T.~Takayanagi,
  ``A Covariant holographic entanglement entropy proposal,''
  JHEP {\bf 0707}, 062 (2007)
  [arXiv:0705.0016 [hep-th]].
  \bibitem{Dong:2016hjy}
  X.~Dong, A.~Lewkowycz and M.~Rangamani,
  ``Deriving covariant holographic entanglement,''
  JHEP {\bf 1611}, 028 (2016)
  [arXiv:1607.07506 [hep-th]].

  \bibitem{Baggioli:2021xuv}
  M.~Baggioli, K.~Y.~Kim, L.~Li and W.~J.~Li,
  Sci. China Phys. Mech. Astron. \textbf{64} (2021) no.7, 270001
  doi:10.1007/s11433-021-1681-8
  [arXiv:2101.01892 [hep-th]].
  
  \bibitem{Horodecki:2009review}
  Horodecki, R., Horodecki, P., Horodecki, M.,  Horodecki, K. (2009).
  ``Quantum entanglement.''
  Reviews of modern physics, 81(2), 865.

  \bibitem{vidal:2002}
  Vidal, G. and Werner, R.F., 2002.
  ``A computable measure of entanglement'',
  Physical Review A, 65(3), p.032314. quant-ph:0102117

  \bibitem{Nishioka:2006gr}
  T.~Nishioka and T.~Takayanagi,
  ``AdS Bubbles, Entropy and Closed String Tachyons,''
  JHEP {\bf 0701}, 090 (2007)
  [hep-th/0611035].

  \bibitem{Klebanov:2007ws}
  I.~R.~Klebanov, D.~Kutasov and A.~Murugan,
  ``Entanglement as a probe of confinement,''
  Nucl.\ Phys.\ B {\bf 796}, 274 (2008)
  [arXiv:0709.2140 [hep-th]].

  \bibitem{Pakman:2008ui}
  A.~Pakman and A.~Parnachev,
  ``Topological Entanglement Entropy and Holography,''
  JHEP {\bf 0807}, 097 (2008)
  [arXiv:0805.1891 [hep-th]].

  \bibitem{Zhang:2016rcm}
  S.~J.~Zhang,
  ``Holographic entanglement entropy close to crossover/phase transition in strongly coupled systems,''
  Nucl.\ Phys.\ B {\bf 916}, 304 (2017)
  [arXiv:1608.03072 [hep-th]].

  \bibitem{Zeng:2016fsb}
  X.~X.~Zeng and L.~F.~Li,
  ``Holographic Phase Transition Probed by Nonlocal Observables,''
  Adv.\ High Energy Phys.\  {\bf 2016}, 6153435 (2016)
  [arXiv:1609.06535 [hep-th]].

  \bibitem{Ling:2015dma}
  Y.~Ling, P.~Liu, C.~Niu, J.~P.~Wu and Z.~Y.~Xian,
  ``Holographic Entanglement Entropy Close to Quantum Phase Transitions,''
  JHEP \textbf{04} (2016), 114
  [arXiv:1502.03661 [hep-th]].

  \bibitem{Ling:2016wyr}
  Y.~Ling, P.~Liu and J.~P.~Wu,
  ``Characterization of Quantum Phase Transition using Holographic Entanglement Entropy,''
  Phys. Rev. D \textbf{93} (2016) no.12, 126004
  [arXiv:1604.04857 [hep-th]].

  \bibitem{Ling:2016dck}
  Y.~Ling, P.~Liu, J.~P.~Wu and Z.~Zhou,
  ``Holographic Metal-Insulator Transition in Higher Derivative Gravity,''
  Phys.\ Lett.\ B {\bf 766}, 41 (2017)
  [arXiv:1606.07866 [hep-th]].

  \bibitem{Kuang:2014kha}
  X.~M.~Kuang, E.~Papantonopoulos, and B.~Wang,
  ``Entanglement Entropy as a Probe of the Proximity Effect in Holographic Superconductors,''
  JHEP {\bf 1405}, 130 (2014)
  [arXiv:1401.5720 [hep-th]].

  \bibitem{Guo:2019vni}
  H.~Guo, X.~M.~Kuang and B.~Wang,
  ``Holographic entanglement entropy and complexity in St\:uckelberg superconductor,''
  Phys. Lett. B \textbf{797} (2019), 134879
  [arXiv:1902.07945 [hep-th]].

  \bibitem{Mahapatra:2019uql}
  S.~Mahapatra,
  ``Interplay between the holographic QCD phase diagram and mutual \& $n$-partite information,''
  JHEP \textbf{04} (2019), 137
  [arXiv:1903.05927 [hep-th]].

  \bibitem{hayden2013holographic}
  P.~Hayden, M.~Headrick, and A.~Maloney, ``Holographic mutual information is
  monogamous,'' Physical Review D, vol.~87, no.~4, p. 046003, 2013.

  \bibitem{Takayanagi:2017knl}
  T.~Takayanagi and K.~Umemoto,
  ``Holographic Entanglement of Purification,''
  arXiv:1708.09393 [hep-th].

  \bibitem{Chu:2019etd}
  J.~Chu, R.~Qi and Y.~Zhou,
  JHEP \textbf{03} (2020), 151
  doi:10.1007/JHEP03(2020)151
  [arXiv:1909.10456 [hep-th]].

  \bibitem{Li:2021rff}
  Y.~Z.~Li, C.~Y.~Zhang and X.~M.~Kuang,
  Sci. China Phys. Mech. Astron. \textbf{64} (2021) no.12, 120413
  doi:10.1007/s11433-021-1791-1
  [arXiv:2102.12171 [hep-th]].

  \bibitem{Liu:2021rks}
  P.~Liu, C.~Niu, Z.~J.~Shi and C.~Y.~Zhang,
  JHEP \textbf{08} (2021), 113
  doi:10.1007/JHEP08(2021)113
  [arXiv:2104.08070 [hep-th]].

  \bibitem{Blake:2016wvh}
  M.~Blake,
  ``Universal Charge Diffusion and the Butterfly Effect in Holographic Theories,''
  Phys.\ Rev.\ Lett.\  {\bf 117}, no. 9, 091601 (2016)
  [arXiv:1603.08510 [hep-th]].

  \bibitem{Blake:2016sud}
  M.~Blake,
  ``Universal Diffusion in Incoherent Black Holes,''
  Phys.\ Rev.\ D {\bf 94}, no. 8, 086014 (2016)
  [arXiv:1604.01754 [hep-th]].

  \bibitem{Ling:2016ibq}
  Y.~Ling, P.~Liu and J.~P.~Wu,
  ``Holographic Butterfly Effect at Quantum Critical Points,''
  JHEP {\bf 1710}, 025 (2017)
  [arXiv:1610.02669 [hep-th]].

  \bibitem{Ling:2016wuy}
  Y.~Ling, P.~Liu and J.~P.~Wu,
  ``Note on the butterfly effect in holographic superconductor models,''
  Phys.\ Lett.\ B {\bf 768}, 288 (2017)
  [arXiv:1610.07146 [hep-th]].

  \bibitem{Israel:1967wq}
  W.~Israel,
  Phys. Rev. \textbf{164} (1967), 1776-1779
  doi:10.1103/PhysRev.164.1776

  \bibitem{Carter:1971zc}
  B.~Carter,
  Phys. Rev. Lett. \textbf{26} (1971), 331-333
  doi:10.1103/PhysRevLett.26.331

  \bibitem{Ruffini:1971bza}
  R.~Ruffini and J.~A.~Wheeler,
  Phys. Today \textbf{24} (1971) no.1, 30
  doi:10.1063/1.3022513



  \bibitem{Volkov:1989fi}
  M.~S.~Volkov and D.~V.~Galtsov,
  JETP Lett. \textbf{50} (1989), 346-350

  \bibitem{Bizon:1990sr}
  P.~Bizon,
  Phys. Rev. Lett. \textbf{64} (1990), 2844-2847
  doi:10.1103/PhysRevLett.64.2844

  \bibitem{Greene:1992fw}
  B.~R.~Greene, S.~D.~Mathur, and C.~M.~O'Neill,

  Phys. Rev. D \textbf{47} (1993), 2242-2259
  doi:10.1103/PhysRevD.47.2242
  [arXiv:hep-th/9211007 [hep-th]].

  \bibitem{Luckock:1986tr}
  H.~Luckock and I.~Moss,
  Phys. Lett. B \textbf{176} (1986), 341-345
  doi:10.1016/0370-2693(86)90175-9

  \bibitem{Bekenstein:1974sf}
  J.~D.~Bekenstein,
  Annals Phys. \textbf{82} (1974), 535-547
  doi:10.1016/0003-4916(74)90124-9

  \bibitem{Bekenstein:1995un}
  J.~D.~Bekenstein,
  Phys. Rev. D \textbf{51} (1995) no.12, R6608
  doi:10.1103/PhysRevD.51.R6608

  \bibitem{Hartnoll:2008kx}
  S.~A.~Hartnoll, C.~P.~Herzog and G.~T.~Horowitz,
  JHEP \textbf{12} (2008), 015
  doi:10.1088/1126-6708/2008/12/015
  [arXiv:0810.1563 [hep-th]].

  \bibitem{Cai:2015cya}
  R.~G.~Cai, L.~Li, L.~F.~Li and R.~Q.~Yang,
  Sci. China Phys. Mech. Astron. \textbf{58} (2015) no.6, 060401
  doi:10.1007/s11433-015-5676-5
  [arXiv:1502.00437 [hep-th]].

  \bibitem{Gonzalez:2013}
  P. A. Gonzalez, Eleftherios Papantonopoulos, Joel Saavedra and Yerko Vasquez,
  ``Four-Dimensional Asymptotically AdS Black Holes with Scalar Hair''
  [arxiv:1309.2161v2 [gr-qc]]

  \bibitem{Martinez:2004}
  Cristian Martinez, Ricardo Troncoso, Jorge Zanelli
  ``Exact black hole solution with a minimally coupled scalar field''
  [arXiv:hep-th/0406111]

  \bibitem{Nishioka:2009un}
  T.~Nishioka, S.~Ryu and T.~Takayanagi,
  J. Phys. A \textbf{42} (2009), 504008
  doi:10.1088/1751-8113/42/50/504008
  [arXiv:0905.0932 [hep-th]].

  \bibitem{Liu:2019qje}
  P.~Liu, Y.~Ling, C.~Niu and J.~P.~Wu,
  JHEP \textbf{09} (2019), 071
  doi:10.1007/JHEP09(2019)071
  [arXiv:1902.02243 [hep-th]].

  \bibitem{Fischler:2012}
  Willy Fischler, Arnab Kundu, Sandipan Kundua,
  ``Holographic Mutual Information at Finite Temperature''
  [arXiv:1212.4764 [hep-th]]

\end{thebibliography}
\end{document}